\pdfoutput=1
\documentclass[useAMS,usenatbib]{mn2e}
\usepackage{color}
\usepackage{soul}
\usepackage[normalem]{ulem}
\usepackage{amsmath}
\usepackage{lscape}
\usepackage{hyperref}
\usepackage[dvips]{graphicx}
\usepackage{pslatex}

\definecolor{red1}{RGB}{198,40,40}
\title[\textit{Fermi}-LAT high $z$ AGN and the EBL]{\textit{Fermi}-LAT high $z$ AGN and the Extragalactic Background Light}
\author[Thomas Armstrong, Anthony M. Brown and Paula M. Chadwick]{Thomas Armstrong$^{1,2}$\thanks{E-mail: thomas.armstrong@physics.ox.ac.uk}, Anthony M.
Brown$^{1}$ and Paula M. Chadwick$^{1}$\\
$^{1}$Centre for Advanced Instrumentation, Department of Physics, Durham University, South Road, Durham, DH1 3LE, UK\\
$^{2}$University of Oxford, Department of Astrophysics, Denys Wilkinson Building, 1 Keble Road, Oxford, OX1 3RH, UK}
\begin{document}

\date{Accepted 2017 May 24}

\pagerange{\pageref{firstpage}--\pageref{lastpage}} \pubyear{2017}

\maketitle

\label{firstpage}

\begin{abstract}

Observations of distant gamma-ray sources are hindered by the presence of the extragalactic background light (EBL). In order to understand the physical processes that result in the observed spectrum of sources, it is imperative that a good understanding of the EBL is included. In this work, an investigation into the imprint of the EBL on the observed spectra of high redshift \textit{Fermi}-LAT AGN is presented. By fitting the spectrum below $\sim$10~GeV, an estimation of the un-absorbed intrinsic source spectrum is obtained; by applying this spectrum to data up to 300~GeV, it is then possible to derive a scaling factor for different EBL models. A second approach uses 5 sources (PKS 0426-380, 4C +55.17, Ton 116, PG 1246+586 and RBS 1432) which were found to exhibit very high energy emission ($E_{\gamma}>100$~GeV).  Through Monte Carlo simulations it is shown that the observation of VHE photons, despite the large distances of these objects, is consistent with current EBL models. Many of these sources would be observable with the upcoming ground based observatory the Cherenkov Telescope Array (CTA), leading to a better understanding of the EBL.
\end{abstract}


\section{Introduction}

\newcommand{\TA}[1]{{\color{red1} \textbf{TA:}\ \texttt{#1}}}
\newcommand{\CO}[2]{{\color{red1} \textbf{#1:}\ \texttt{#2}}}
\newcommand{\SO}[1]{{\color{red1} \sout{#1}}}

A large fraction of the sources that have so far been detected in the very high energy (VHE, E$_{\gamma}>100$~GeV) regime are blazars \citep{tevcat}. These active galactic nuclei (AGN) have relativistic jets orientated in such a way that the emission is beamed towards Earth, providing a large boost to the observed flux.  Despite this, our ability to detect blazars at energies greater than some 10's to 100's of GeV is hindered by the existence of the Extragalactic Background Light (EBL), which attenuates gamma-rays by way of pair production with lower energy photons (\cite{Nikishov1962}, \cite{1967PhRv..155.1404G}, \cite{1967PhRv..155.1408G} and \cite{2013APh....43..112D} for a recent review). 

The EBL is thought to consist of UV-FIR (far infrared) radiation from stars (with UV-optical representing the main attenuation band for gamma-rays at energies below 100~GeV), either produced directly from their surfaces or via reprocessing by dust within their host galaxies, with an additional component coming from optically bright AGN \citep{fermiEBL}. The evolution of the EBL density is of considerable interest as it probes models of galaxy and star formation/evolution. However, direct measurements are difficult due to the presence of foreground zodiacal and Galactic light \citep{2001ARA&A..39..249H}. Additionally, measurements of the EBL in the local universe ($z=0$) provide little information about the evolution of the EBL through past epochs. This evolution is especially important when considering the attenuation of gamma-rays from distant sources and its understanding therefore represents one of the major science goals of the \textit{Fermi}-LAT space based instrument, which is able to observe gamma-rays from 100~MeV to above 300~GeV \citep{fermi}.

For AGN at a redshift of \textit{z}=1, the attenuation from the EBL quickly becomes significant above 10~GeV, therefore the \textit{Fermi}-LAT instrument is well suited to observe, and place constraints on, the effect of the EBL. By modelling the un-absorbed part of a given AGN spectrum, it is possible to obtain an indication of the intrinsic source spectrum. Combining this with different EBL models, the amount of EBL absorption, and therefore density, can be inferred by evaluating the magnitude of the softening of the spectrum above $\sim$10~GeV. This method has been adopted to derive EBL limits by several authors (\cite{fermiEBL}, \cite{2012Sci...338.1190A} \cite{2013A&A...550A...4H}, \cite{2015ApJ...812...60B}, \cite{2016arXiv160901095M}, \cite{2016A&A...590A..24A} and \cite{2016arXiv161009633M}).

An alternative method which can be used to set limits on the EBL is to measure the maximum observed photon energy from distant gamma-ray emitters, such as AGN or gamma-ray bursts (GRBs). The observation of high energy photons from large redshifts, where the universe is thought to be optically thick to gamma-rays, can challenge current EBL models.  Currently, the most distant AGN detected in the VHE band are the gravitationally-lensed blazar B0 0218+357 ($z=0.94$) \citep{2016arXiv160901095M} and the FSRQ PKS 1441+25 ($z=0.939$)  \citep{Ahnen2015}, both detected with the MAGIC telescope, with the latter also detected by VERITAS \citep{2015arXiv150807251B}. Additionally, the second \textit{Fermi}-LAT catalogue of hard sources (50~GeV - 2~TeV) contains a large sample of 271 AGN with a redshift range out to 2.1 \citep{2fhl}. The two GRBs, GRB 090902B ($z=1.822$) \citep{GRB090902B} and GRB 08916C ($z=4.35$) \citep{GRB080916C}, have also aided in providing limits to the EBL density.

In an attempt to increase the sample of VHE AGN, previous work (\cite{Arm2015}, \cite{Arm2016}) has focused on developing clustering methods to identify extragalactic sources with $E_{\gamma} > 100$~GeV in the \textit{Fermi}-LAT data set.  The chosen algorithm, \textsc{dbscan}, was first proposed in \cite{dbscan} and was designed to efficiently detect clusters of arbitrary shape in noisy data sets. The algorithm works by evaluating the number of events around a detected photon within the data set; if these exceed a given number then they are added to a cluster and in turn evaluated, accumulating further events to the cluster that satisfy the same condition\footnote{For the \textsc{dbscan} sources presented in this work that were found in \cite{Arm2016}, the search radius around a given point was 0.4 degrees, the PSF at 100~GeV, and the number of required events started at 2 and scaled upward with the integrated Galactic background emission}. The \textsc{dbscan} algorithm makes no assumption regarding the underlying source spectrum. The ability to identify clusters within sparse noisy data sets made \textsc{dbscan} an ideal choice for analysing \textit{Fermi}-LAT VHE data. So far this method has proven to be successful, discovering emission with $E_{\gamma} > 100$~GeV from 45 sources. The sources detected are all either AGN or unassociated sources. One of the aims of this work is to draw attention to six high redshift AGN, PKS 0426-380 ($z=1.11$), 4C +55.17 ($z=0.899$), Ton 116 ($z=1.088$), PG 1246+586 ($z=0.857$), RBS 1432 ($z=1.508$) and TXS 1452+516\footnote{The Blazar TXS 1452+516 ($z=1.522$) was found in a separate work [in prep], using the same method but for $50~\textrm{GeV} > E_{\gamma}>2$~TeV.}, that were found using this method and could have implications for EBL limits. 

This paper is organised as follows, in Section \ref{S:EBL} a brief overview of the EBL and EBL models used in this work will be given. In Section \ref{S:srcInfo} a description of each source of interest, with a focus on the reliability of their redshift determination, will be presented. Section \ref{S:data} will describe the data selection process and Sections \ref{S:specAn} and \ref{S:VHEphoton} will discuss the spectral and `highest energy photon' analyses respectively. Finally Section \ref{S:conc} will conclude and summarise the results presented. 
 %Measurements of the spectra of AGN can provide essential information on the density of the EBL and a sample extending over a large redshift range can reveal the evolution of the EBL on cosmological scales.

%\textcolor{red1}{\section{DBSCAN}}
%\label{S:dbscan}
%
%\textcolor{red1}{ }
%

\section{EBL Models}
\label{S:EBL}

Since direct measurements of the EBL are hindered due to the presence of foreground emission, and because this reveals little about the evolution on cosmological scales, it is necessary to use complex models to estimate the density of EBL photons and therefore the absorption of gamma-rays. There is currently a wide range of models available which adopt different methods to determine the evolution of the EBL density as a function of redshift. 

In this work, five models have been considered: \cite{Gilmore2012} (GIL12) which uses a semi-analytic model of the star formation rate, initial mass function and dust extinction, \cite{Fink2010} (FIN10) and \cite{Kneiske2010} (K\&D10) which use forward evolution models based on observations of \textit{Spitzer, ISO, Hubble, COBE, BLAST} and \textit{GALEX} from which the cosmic star formation rate is inferred, and finally \cite{FRAN08} (FRA08) and \cite{Dominguez11} (DOM11) which use backward evolution models to model the redshift evolution of the luminosity function of galaxies based on number counts. For an overview of the different model types, see \cite{2013APh....43..112D}.

The absorption of distant gamma-rays depends on the optical depth of the EBL, expressed as  $\tau(E,z,n)$, which is dependent on the gamma-ray energy ($E$), the distance ($z$) and the density ($n$) of EBL photons, the last being defined by the choice of model. The total is based on an integral along the line of sight to the target source. For an individual source, the spectrum is therefore attenuated as a function of its energy such that 

\begin{equation}
\frac{dN}{dE}_{obs}=e^{-\tau (E,z,n)} \frac{dN}{dE}_{int}
\label{E:eblspec}
\end{equation}

where $dN/dE_{obs}$ is the observed spectrum and $dN/dE_{int}$ is the intrinsic/unabsorbed spectrum. For the models used in this work, tabulated data of $\tau$ as a function of energy, available for a range of redshift values, were obtained from online resources\footnote{\url{http://www.phy.ohiou.edu/~finke/EBL/} for \cite{Fink2010}, \url{http://www.ias.u-psud.fr/irgalaxies/cib.php} for \cite{Kneiske2010}, \url{http://physics.ucsc.edu/~joel/EBLdata-Gilmore2012} for \cite{Gilmore2012} and \url{http://www.astro.unipd.it/background/} for \cite{FRAN08}. Data from Dominguez were obtained from \url{https://github.com/me-manu/eblstud/blob/master/ebl/ebl_model_files/tau_dominguez10.dat} due to the inactivity of the original link in \cite{Dominguez11}.}. By interpolating these data sets, a function for determining $\tau$ based on the energy and redshift is obtained.

\section{Source Information}
\label{S:srcInfo}
An overview of each source analysed in this work is given here, with a focus on determining the most reliable redshift from literature. The sources are divided into two categories: 1) DBSCAN VHE sources, which were identified in \cite{Arm2016} to have significant emission in the 100~GeV - 3~TeV energy range and 2) a sample of 10 2FHL objects with the highest quoted redshifts within the catalogue which also coincided with the redshift range of the first sample. Two sources where however excluded from the selection, MG4 J000800+4712 which is quoted to have a redshift of 2.1 but is found with \textit{z}=0.28 in all other literature, and PKS 0823-223 due to its proximity to the Galactic plane. The second category of sources was included in order to obtain a more reliable limit for the EBL when considering the fit to the intrinsic and absorbed spectra (see Section \ref{S:specAn}).

\begin{table*}
\centering
\begin{tabular}{lcccccr}
Source & Type & RA & Dec & $z$ & $z_{err}$ & $z_{ref}$ \\ \hline
PKS 0426-380 & BLL & 67.18 & 37.93 & 1.111 & ... & \cite{heidt} \\
4C +55.17 & FSRQ & 149.41 & 55.38 & 0.901 & 0.00019 & \cite{sdssdr13} \\
Ton 116 & BLL & 190.80 & 36.48 & 1.066 & 0.00150 & \cite{sdssdr7} \\
PG 1246+586 &  BLL & 192.08 & 58.34 & 0.847 & 0.00169 & \cite{sdssdr7} \\
RBS 1432 & BLL & 221.50  & 36.35 & 1.565 & +0.275-0.125 & \cite{Richards} \\
TXS 1452+516 & BLL & 223.61 & 51.41 & 1.522 & 0.00183 & \cite{sdssdr12} \\
%PKS 0823-223 & BLL & 126.51 & -22.51 & 0.9103 & 0.0005 & \cite{Falomo1990} \\
GB6 J0043+3426 & FSRQ & 10.95 & 34.44 & 0.966 & ... & \cite{shaw12} \\
B0218+357 & FSRQ & 35.27 & 35.94 & 0.960 & ... & \cite{Lawrence1996} \\
PKS 0235+164 &  BLL & 39.66 & 16.62 & 0.940 & ... & \cite{2massbll} \\
MG2 J043337+2905 & BLL & 68.41 & 29.10 & 0.970 & ... & \cite{bzcat} \\
PKS 0454-234 & FSRQ & 74.26 & -23.41 & 1.003 & ... & \cite{Stickel1989} \\
PKS 0537-441 & BLL & 84.70 & -44.09 & 0.892 & ... & \cite{peterson} \\
TXS 0628-240 & BLL & 97.75 & -24.11 & 1.6 & +0.10-0.05 & \cite{photometryz} \\
OJ 014 & BLL & 122.86 & 1.78 & 1.148 & ... & \cite{esonewz} \\
PKS B1424-418 & FSRQ & 216.98 & -42.11 & 1.522 & 0.002 & \cite{southernZ} \\ 
B2 2114+33 & BLL & 319.06 & 33.66 & 1.596 & ... & \cite{shaw} \\\hline
\end{tabular}
\caption{Summary table of source data used in this study.}
\label{T:srcsum}
\end{table*}

\subsection{DBSCAN VHE Sources}
\label{SS:DBSCANsrc}

\begin{enumerate}
\item[\textbf{PKS 0426-380:}]
Classed as a BL Lac in the 3FGL and situated in the southern hemisphere at RA=67.18$^{\circ}$, Dec=-37.93$^{\circ}$. The 2FHL redshift is quoted as $z=1.111$ as found in \cite{heidt}, where the existence of a closer source at $z=0.559$ implies micro-lensing may be present. This source has previously been noted for producing VHE photons in \cite{Tanaka2013} and \cite{Neronov2015}, which both use Pass 7 data. The new Pass 8 data have reclassified the energy of these photons. 
\item[\textbf{4C +55.17:}] 
Defined as a FSRQ, and located in the northern hemisphere at RA=149.41$^{\circ}$, Dec=55.38$^{\circ}$, in the 2FHL the redshift is quoted as 0.899, most likely obtained from \cite{improvedsdss}. However, more recent measurements from the 13th SDSS data release \citep{sdssdr13} finds a redshift of 0.901$\pm$0.00019 which will be adopted in this work. This source was considered in \cite{Neronov2015} where a redshift of 0.8955 was adopted.
\item[\textbf{Ton 116:}]
A BL Lac located in the northern hemisphere at RA=190.80$^{\circ}$, Dec=36.46$^{\circ}$ with a redshift of 0.0 in the 2FHL indicating that no redshift was available at the time or that it was determined to be unreliable. However, measurements from SDSS indicate a redshift of 1.066$\pm$0.00150 (data release 7, \citep{sdssdr7}) or 1.182$\pm$0.00132 (data release 13 \citep{sdssdr13}); both have data flags associated with the spectrum. This source was also considered in \cite{Neronov2015} in which a redshift of 1.065 was used. Our work adopts the SDSS data release 7 value, which is the same within errors. 
\item[\textbf{PG 1246+586}]
A BL Lac located in the northern hemisphere at RA=192.08$^{\circ}$, Dec=58.34$^{\circ}$ and with an unknown or uncertain redshift in the 2FHL. Measurements from SDSS (data release 7 \citep{sdssdr7}) show a redshift of 0.847$\pm$0.00169. This source was considered in \cite{Neronov2015} who applied the same redshift as used here.
\item[\textbf{RBS 1432:}]
A northern hemisphere BL Lac at RA=221.50$^{\circ}$, Dec=36.35$^{\circ}$ for which the 2FHL again defines the redshift as unknown or uncertain. The redshift in the literature is generally accepted as 1.565 which originates from \cite{Richards}, although the redshift is quoted as being between 1.440 and 1.840.
\item[\textbf{TXS 1452+516:}]
This source was found in a comparative study of DBSCAN and the 2FHL [paper in prep], using the same energy range and observational period. As it was not found to be significant in VHE range (E$_{\gamma}>$100~GeV) range, as with the rest of this sample, we do not claim in to be a VHE source. However as it is not in the 2FHL (but is in the 3FGL) it is included in this list. It is a BL Lac located in the northern hemisphere at RA=223.61$^{\circ}$, Dec=51.41$^{\circ}$ and has a redshift of 1.522$\pm$0.00183, which originates from the 12th SDSS data release \citep{sdssdr12}.

\end{enumerate}

\subsection{High-$z$ 2FHL Sample}
\label{SS:highz2FHL}

\begin{enumerate}
\item[\textbf{GB6 J0043+3426:}]
A northern hemisphere FSRQ at RA=10.95$^{\circ}$, Dec=34.44$^{\circ}$ with a redshift of 0.966 based on observations at the W. M. Keck Observatory and presented in \cite{shaw12}.
\item[\textbf{B0218+357:}]
A northern hemisphere FSRQ at RA=35.27$^{\circ}$, Dec=35.94$^{\circ}$ which is thought to be a gravitationally lensed object with two images of the FSRQ at a redshift of 0.960 \citep{Lawrence1996} and a lensing source at 0.685 \citep{glensz1}. This FSRQ has been observed with MAGIC \citep{2016arXiv160901095M} and was used to determine an EBL correction factor.% of $0.99\pm0.34_{\textrm{stat}} ~^{+0.15} _{-0.18, \textrm{sys}} \cdot \tau_{\textrm{GIL12}}$.
\item[\textbf{PKS 0235+164:}]
A northern hemisphere BL Lac at RA=39.66$^{\circ}$, Dec=16.62$^{\circ}$ which has an associated redshift of 0.940 as stated in \cite{2massbll}. 
\item[\textbf{MG2 J043337+2905:}]
A northern hemisphere BL Lac at RA=68.41$^{\circ}$, Dec=29.10$^{\circ}$ with a redshift of 0.970 found in \cite{bzcat}.
\item[\textbf{PKS 0454-234:}]
A southern hemisphere FSRQ at RA=74.26$^{\circ}$, Dec=-23.41$^{\circ}$. The redshift is given as 1.003 based on absorption lines in \cite{Stickel1989}. 
\item[\textbf{PKS 0537-441:}]
A southern hemisphere BL Lac at RA=84.71$^{\circ}$, Dec=-44.09$^{\circ}$ with a redshift of 0.892 \citep{peterson}.
\item[\textbf{TXS 0628-240:}]
A southern hemisphere BL Lac at RA=97.75$^{\circ}$, Dec=-24.11$^{\circ}$ which has an associated photometric redshift of 1.6$^{+0.10}_{-0.05}$ \citep{photometryz} or limits of 1.239 to 1.91 based on spectroscopic measurements \citep{shaw}. In this work we use the photometric redshift. 
\item[\textbf{OJ 014:}]
A southern hemisphere BL Lac at RA=122.86$^{\circ}$, Dec=1.78$^{\circ}$ with a redshift of 1.148 from \cite{esonewz}.
\item[\textbf{PKS B1424-418:}]
A southern hemisphere FSRQ at RA=216.98$^{\circ}$, Dec=-42.11$^{\circ}$ with redshift 1.522$\pm$0.002 from \cite{southernZ}.
\item[\textbf{B2 2114+33:}]
A southern hemisphere BL Lac located at RA=319.06$^{\circ}$, Dec=33.66$^{\circ}$. A redshift is given as 1.596 based on significant broad emission feature identified with C$_{\textrm{IV}}$ consistent with a weak bump in the far blue at Ly$\alpha$.  It is suggested that if the emission feature is false that this would then at least represent a lower limit \citep{shaw}. 
\end{enumerate}

A summary of all these sources and their basic data can be found in Table \ref{T:srcsum}.

\section{Data Selection}
\label{S:data}

Each source presented in this work was evaluated using the Pass 8 processed data \citep{pass8},  which provides several improvements over the previous Pass 7 reprocessed data set \citep{pass7rep}\footnote{Pass 7 was previously used to evaluate some of the sources presented in this work (\cite{Tanaka2013} and \cite{Neronov2015}).}. These include better energy and angular resolution, increased effective area, an extended observable energy range (around 10~MeV \textless~$E_{\gamma}$ \textless 3~TeV) and better background characterisation, resulting in improved point source sensitivity. As a result of this re-evaluation of the data, the energies of many events have been updated. 

To investigate each source, 8 years of data ranging from August 2008 to September 2016 (MET  239557417 - 494930839) and with energies from 100~MeV to 3~TeV were downloaded from the \textit{Fermi} data servers\footnote{http://fermi.gsfc.nasa.gov/ssc/data/}. All source class events were retained for both front and back converting photons. Additionally, the recommended filter expression \textit{`DATA\_QUAL$>$0 \&\& LAT\_CONFIG==1'} was applied in order to remove any sub-optimal data affected by spacecraft events. Finally a zenith cut of 90$^{\circ}$ was implemented in order to remove any $\gamma$-rays induced by cosmic-ray interactions in the Earth's atmosphere.

\section{Spectral Analysis}
\label{S:specAn}
To investigate the effect of the EBL absorption on the spectrum of each source presented in Section \ref{S:srcInfo}, a binned likelihood analysis was performed. An initial fit was obtained for each source below a threshold energy $E_{max}$, here defined to be the point where there is less than 0.1\% of emission attenuated by the EBL, in order to represent the intrinsic spectrum. For a source at redshift $\sim$1 this corresponds to an energy of $\sim$10~GeV. The values for each source can be found in Table \ref{T:SPEC}. 

From each source data set comprising a 10$^{\circ}$ region of interest (ROI) around the source, an initial model is constructed from the 3FGL which includes sources out to a further 10$^{\circ}$ fixed to the 3FGL values. The normalisations and spectral indices of the sources within the ROI were allowed to vary. Also included were the extragalactic diffuse emission model (iso\_P8R2\_SOURCE\_V6\_v06.txt) with a free normalisation, and the Galactic diffuse template (gll\_iem\_v06.fits), which was multiplied by a power law in energy and the normalisation of which was free to vary. The sources of interest were modelled with both a power law and a log parabola in the form

\begin{equation}
\frac{dN}{dE}=N_{0} \left(\frac{E}{E_0}\right)^{\Gamma}, ~~~~~~~ \frac{dN}{dE}=N_{0} \left( \frac{E}{E_0}\right) ^{-(a+b~\textrm{log}(E/E_0))}
\label{E:pwllgp}
\end{equation}

where $N_0$ is the normalisation, $\Gamma$ is the spectral index of the power law, $a$ is the log parabola index, $b$ is the curvature and $E_0$ is a scaling factor. In the analysis of each source, all parameters excluding $E_{0}$ were left free to vary and a full binned likelihood analysis was performed for each ROI, returning the best fit model. The best fit intrinsic model parameters can be found in Table \ref{T:SPEC}. 

\begin{table}
\centering
\resizebox{\columnwidth}{!}{
\begin{tabular}{lcccc}
 			&	$N_{0}$ $\times 10^{-12}$			 & $\Gamma$  & ($a,b$) &$E_{max}$   \\
Source &  ph~cm$^{-2}~$s$^{-1}~$MeV$^{-1}$ &  (PL) & 			 (LGP) & (GeV) \\ \hline
PKS 0426-380 			& 	54.09 	& -1.99 	&  	1.99  	&  9.00 \\%LGP
								&				&			&		0.06		&  \\ 
4C +55.17 				&  10.64	& -1.90	&		1.92		&	10.58 \\% LGP
								&				&			&		0.06		&  \\ 
Ton 116 					&  1.57		&	-1.62	&		1.64		&	 9.42 \\  % PL \hline
								&				&			&		-0.02		&  \\ 
PG 1246+586 			&  4.27		&	-1.81	&		1.82		&	 11.20 \\ % PL \hline
								&				&			&		-0.02		&  \\ 
RBS 1432 					& 1.21		&	-1.69	&		1.64		&	 6.96\\ % PL \hline
								&				&			&		0.07		&  \\ 
TXS 1452+516 			& 3.3 		&	-1.98	&		1.99		&	 6.89\\  % PL \hline
								&				&			&		0.02		&  \\ 
%PKS 0823-223 			&  4.20	&	-1.89	&						&	 15.06\\  % PL \hline
GB6 J0043+3426		& 2.62		&	-1.90	&		1.91		&	 9.94\\  % PL \hline
								&				&			&		-0.03		&  \\ 
B0218+357 				& 10.25		&	-2.25	&		2.29		&	 10.13\\  % PL \hline
								&				&			&		0.04		&  \\ 
PKS 0235+164 			&  14.80	&	-2.08	&		2.12		&	 10.17\\ % LGP
								&				&			&		0.07		& \\ 
MG2 J043337+2905 	&  3.26		&	-2.00	&		1.99		&	 9.91\\ % PL \hline
								&				&			&		0.03		&  \\ 
PKS 0454-234 			& 	27.09	&	-2.12	&		2.19		& 9.65	 \\ % LGP
								&				&			&		0.08		& \\ 
PKS 0537-441 			&  23.95	&	-2.03	&		2.07		&	 10.66\\% LGP
								&				&			&		0.05		&\\ 
TXS 0628-240 			& 	3.23		&	-1.73	&		1.72		&	 8.12\\  % PL \hline
								&				&			&		0.02		&  \\ 
OJ 014 			 			&  3.84		&	-2.00	&		2.01		&	 8.75\\ % PL \hline
								&				&			&		0.04		&  \\ 
PKS B1424-418 		&  47.58	&	-2.09	&		2.15		&	 6.75\\ % LGP
								&				&			&		0.05	&\\
B2 2114+33 				&  2.42		&	-1.64 &		1.57		&	 6.56\\ % PL \hline
								&				&			&		0.10		&  \\  \hline
\end{tabular}
}
\caption{Intrinsic model parameters for each source derived from the binned likelihood analysis below $E_{max}$. The parameters correspond to those found in equation \ref{E:pwllgp}. PL = power law model and LGP = log-parabola model.}
\label{T:SPEC}
\end{table}

In order to investigate the level of EBL absorption present, the initial fit was scaled with $e^{-\alpha \cdot \tau (E,z,n)}$, as in Equation \ref{E:eblspec}, where the scaling factor $\alpha$ has now been introduced (here a value of $\alpha=1$ would return the original EBL model absorption level). By taking the scaled spectral model, and fitting this to to data between 100~MeV and 300~GeV, which includes both the absorbed and unabsorbed sections of the spectrum, it is possible to scan through values of $\alpha$ and identify the optimal level of EBL absorption (as was performed in \cite{2012Sci...338.1190A} and \cite{2013A&A...550A...4H}; a more detailed description of the process is given in Appendix \ref{A:1}).

\begin{figure*}
\centering
\includegraphics[width=0.89\textwidth]{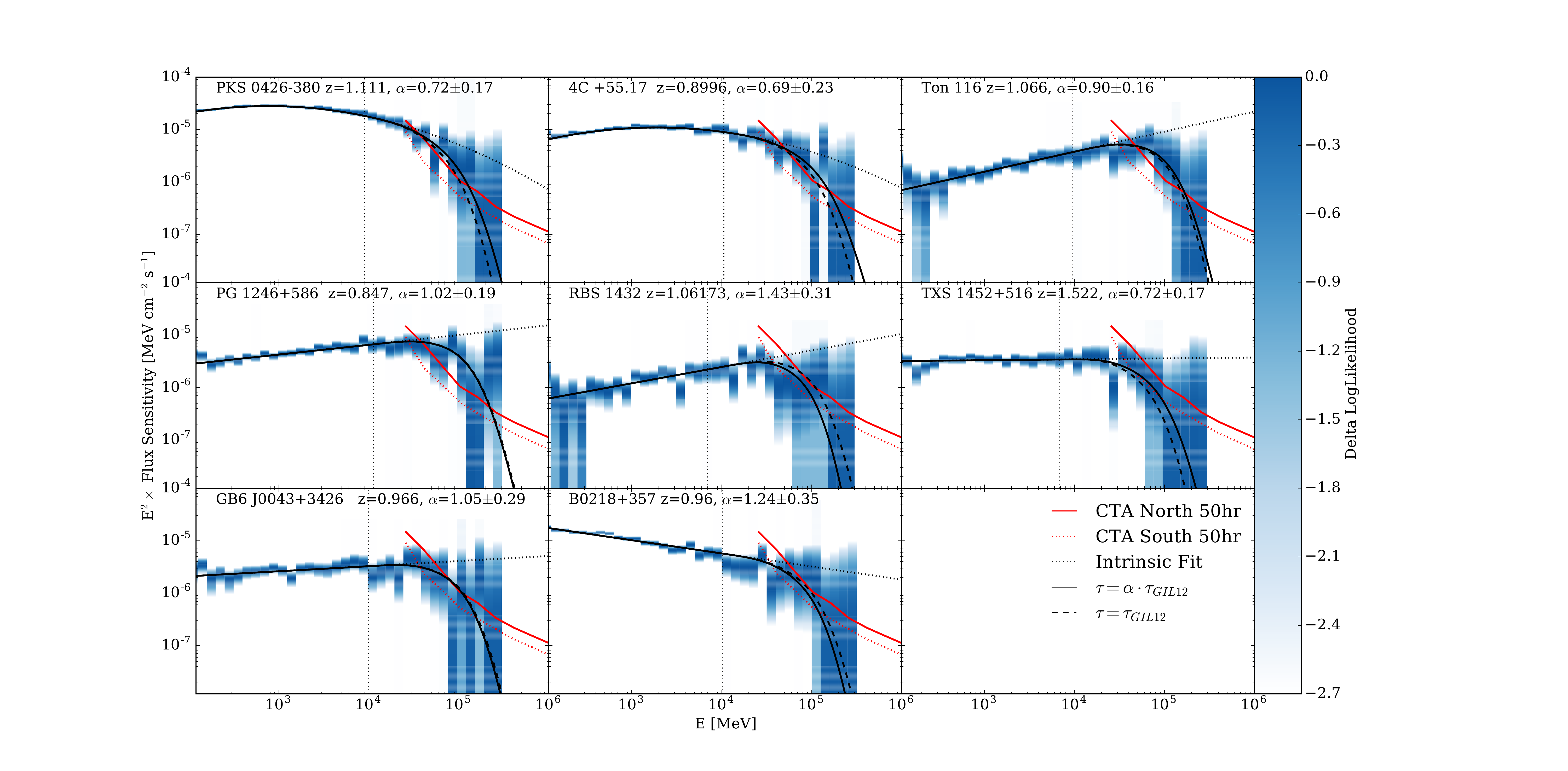}
\caption{Spectral  energy distribution for each source between 100~MeV and 300~GeV. For each energy bin the delta log likelihood determined from the binned likelihood analysis is plotted, where all but the normalisation is fixed to the best fit intrinsic model. The vertical dotted line shows the boundary between the intrinsic and absorbed spectrum. Also shown are the intrinsic spectrum fit (black dotted line), the intrinsic model including EBL absorption, pure model (GIL12, black dashed line) and best fit modification to GIL12 (black solid line). Lastly, as an indication of whether these sources would be observable by future ground-based gamma-ray observatories, the predicted sensitivity of the future CTA observatory is shown (\url{https://www.cta-observatory.org/science/cta-performance/}).}
\label{F:sed1}
\end{figure*}

\begin{figure*}
\centering
\includegraphics[width=0.89\textwidth]{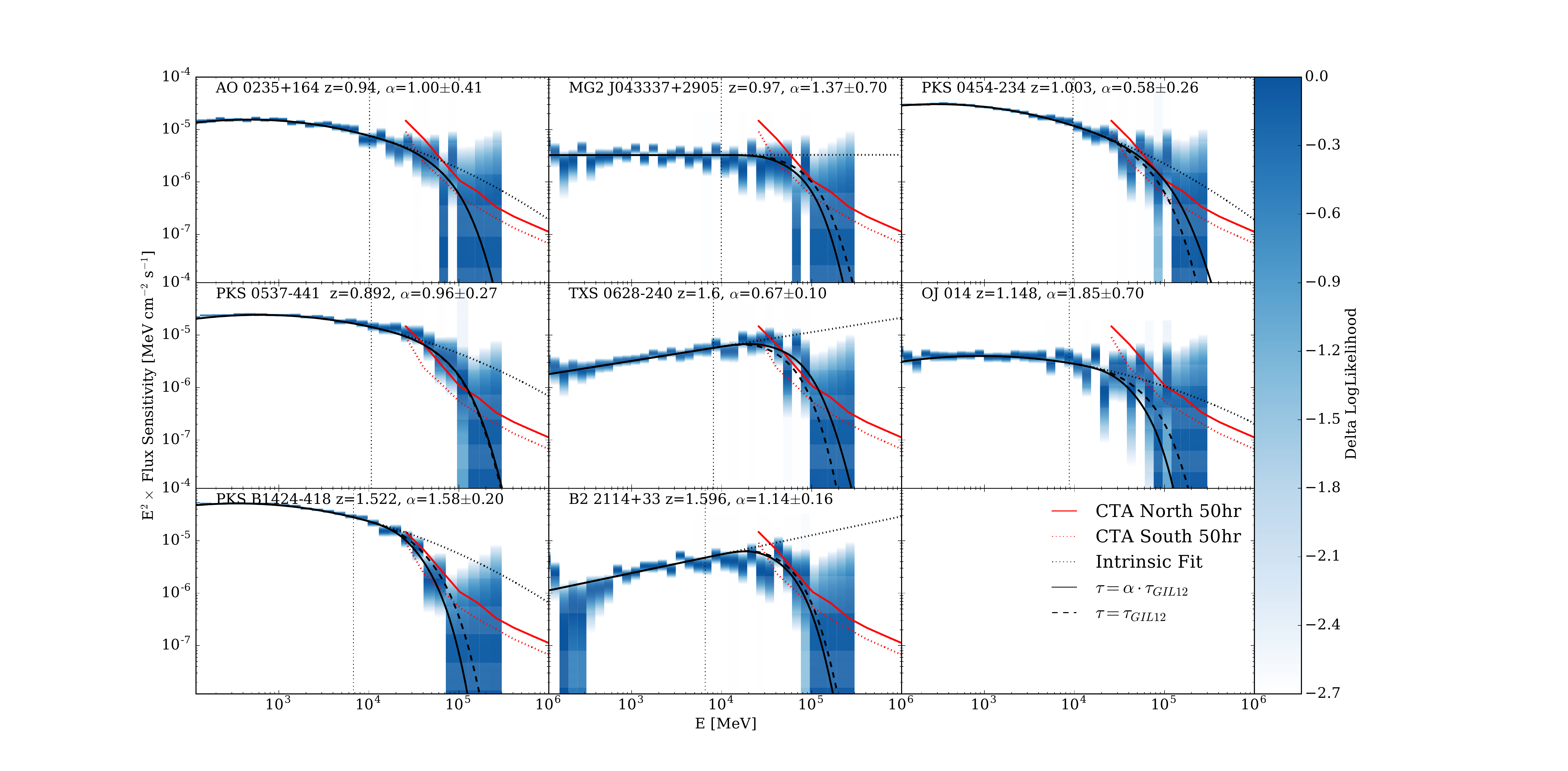}
\caption{See Figure \ref{F:sed1} caption.}
\label{F:sed2}
\end{figure*}

For each EBL scaling factor, a likelihood was generated for the attenuated power law and log parabola models. The latter was chosen if the Test Statistic (TS) between the two models, with the maximum likelihood value of $\alpha$ applied, was greater than 16 (the power law is the null hypothesis, i.e. TS=2log$[\mathcal{L}_{\textrm{max}}(Log Parabola)/\mathcal{L}_{\textrm{max}}(Power Law)]$)\footnote{This is the value used by the \textit{Fermi}-LAT collaboration in the 3FGL to chose a given spectral model over a power law \cite{2015ApJS..218...23A}.}. In Figures \ref{F:sed1} and \ref{F:sed2} the delta logLikelihood can be seen for each energy bin of the full spectral energy distribution along with the intrinsic spectrum (black dotted line) which is calculated based on events with energies below that indicated by the vertical dotted line. The black dashed line shows the absorbed spectrum calculated using the chosen EBL model (in this case GIL12) and the solid black line the resulting the best maximum likelihood fit when GIL12 is modified by a scaling factor $\alpha$. The individual EBL scaling factors for each source are given in the Figures.

\begin{figure}
\centering
\includegraphics[width=0.5\textwidth]{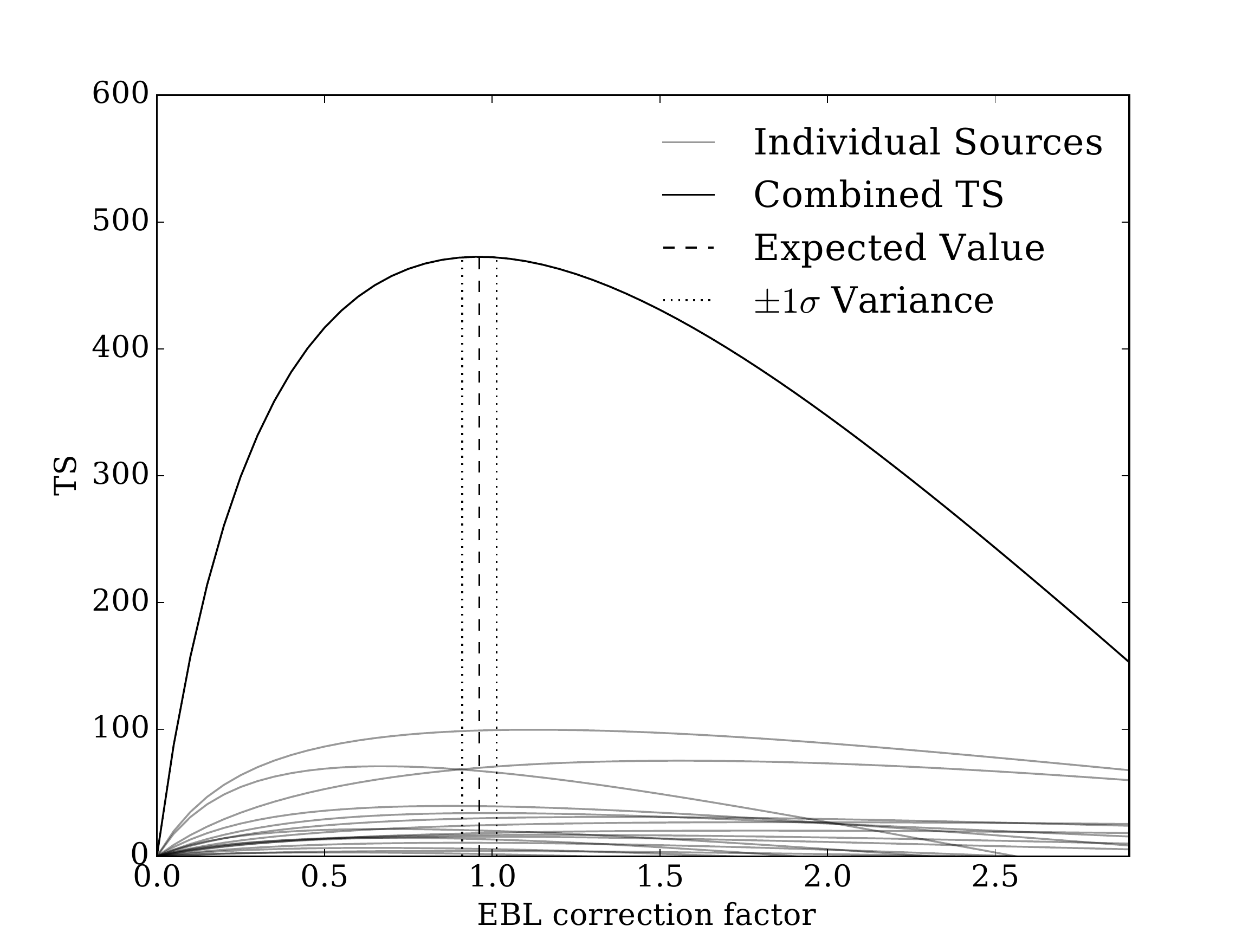}
\caption{Combined TS (solid black) from each individual source (grey lines) as a function of EBL scaling factor. This is based on the GIL12 model for which the best fit EBL scaling factor corresponds to $\alpha=0.96\pm0.05$ where the error is the $1\sigma$ standard deviation. }
\label{F:alphascan}
\end{figure}

\begin{table}
\centering
\begin{tabular}{lcc}
		 & \multicolumn{2}{c}{EBL correction Factor} \\ 
Model 						& Best Fit 	& LogParabola \\ \hline
\citep{Fink2010} 		& 1.31$\pm$0.08 & 1.24$\pm$0.08\\
 \citep{Kneiske2010} & 1.31$\pm$0.07 & 1.24$\pm$0.07\\
 \citep{Gilmore2012} 	& 0.95$\pm$0.05 & 0.90$\pm$0.05 \\
 \citep{Dominguez11} & 1.85$\pm$0.11 & 1.75$\pm$0.11 \\
 \citep{FRAN08} 		& 1.85$\pm$0.11 & 1.71$\pm$0.11\\ \hline
\end{tabular}
\caption{Derived EBL correction factor for each of the 5 models used in this work. The `Best Fit' column is based on whichever of the power law or log parabola models is a better fit for each source, while the last column uses only log parabola spectral models for every source, regardless of whether this is the best fit model.}
\label{T:eblcorr}
\end{table}

From the likelihood distribution for each source as a function of the EBL correction factor, the TS was calculated with respect to the null hypothesis that there is no absorption from the EBL (i.e. TS=2log$[\mathcal{L}(\alpha)/\mathcal{L}(\alpha=0)]$). These distributions were then summed in order to obtain a combined TS distribution, allowing an overall EBL correction to be calculated. The result of this process applied to the GIL12 model can be seen in Figure \ref{F:alphascan} from which an optimal correction factor of $\alpha_{\textrm{GIL12}}=0.95\pm0.05$ was derived (1 standard deviation uncertainty).  The EBL normalisation for the other models gives $\alpha_{\textrm{K\&D10}}=1.31\pm0.07$, $\alpha_{\textrm{FIN10}}=1.31\pm0.08$, $\alpha_{\textrm{DOM11}}=1.85\pm0.11$ and $\alpha_{\textrm{FRA08}}=1.85\pm0.11$ for the redshift range $0.897<z<1.596$. It is clear that the GIL12 model is consistent with the results, while it is suggested that the other models underestimate the EBL density. 

As noted above, the quoted errors are statistical only. One of the main sources of systematic error in this analysis comes from the choice of spectral model, where it is clear that an incorrect choice will strongly affect the derived EBL scaling factor. This was addressed by \cite{2012Sci...338.1190A} who concluded that the assumption that spectral cutoffs existed in the EBL absorption energy range was acceptable. However, \cite{2012Sci...338.1190A} assumed each source to be best represented by a log parabola. We have therefore repeated our analysis using only log parabola models, which provides a lower limit for the EBL correction factor. This leads to the following derived scaling factors: $\alpha_{\textrm{GIL12}}=0.90\pm0.05$, $\alpha_{\textrm{K\&D10}}=1.24\pm0.07$, $\alpha_{\textrm{FIN10}}=1.24\pm0.08$, $\alpha_{\textrm{FRA08}}=1.71\pm0.11$ and $\alpha_{\textrm{DOM11}}=1.75\pm0.11$. The derived EBL scaling factors are summarised in Table \ref{T:eblcorr}.

In each panel of Figures \ref{F:sed1} and \ref{F:sed2}, the expected sensitivity for 50 hours of observation of the two arrays of the future ground based gamma-ray observatory, CTA, are shown (publicly available at \url{https://www.cta-observatory.org/science/cta-performance/}, see \cite{2013APh....43..171B} for description of method). Since CTA will possess a much improved sensitivity compared to current telescopes, it will be well placed to develop further our understanding of the EBL by observing AGN in the range $0<z<1$ and at higher energies than is possible with \textit{Fermi}. This work shows that several of our sources should be detectable with 50 hours of observation, while others may also be detectable with deeper observations. Further observations with CTA would aid in reducing systematics in the choice of best fit spectral model.

\section{VHE Photons}
\label{S:VHEphoton}

An alternative method for probing the EBL is to identify the highest energy photons associated with distant AGN. Here, detection of a photon from a high-$z$ source at energies where the density of the EBL is thought to attenuate the signal almost entirely can bring EBL models into question. In this section, we investigate whether this is the case for the sources presented in the DBSCAN VHE sample which have already exhibited significant emission above 100~GeV (which therefore excludes TXS 1452+516), despite their large redshifts.   

For this analysis, it is particularly important to ensure that the VHE photons are genuinely associated with the AGN. To verify this, a subset of data was selected for each source covering energies above 100~GeV only for a 1$^{\circ}$ region around the source position (much larger than the PSF at these energies). The \textit{Fermi} science tool \textsc{gtsrcprob} was applied to this dataset in order to calculate the probability that each photon is associated with a source in or around the ROI. This is based on the likelihood of each photon and convolves a source model with the instrument response functions. For the source model, the best fit derived in the previous section without EBL absorption was used. Additionally, in order to account for diffuse components such as the Galactic diffuse model, the \textit{Fermi} science tool \textsc{gtdiffrsp} was applied in order to add a diffuse response to the input data. The energies of the highest energy photons and the probabilities of their being associated with the source in question can be seen in Table \ref{T:HEPprob}. It is worth noting that, if the attenuation from the EBL is included, these probabilities will reduce. For example, the highest energy photon probability associated with PKS 0426-380 drops from 0.99983 to 0.99899 when the GIL12 model is included. For out purpouse we assume this to be a negligible difference.

\begin{table}
\centering
\resizebox{\columnwidth}{!}{
\begin{tabular}{lcccccc}
source & E1 & P(E1) & $\sigma$(E1) & E2 & P(E2) & $\sigma$(E2) \\ \hline
PKS 0426-380 & 122.0 & 0.99983 & 3.76 & 115.8 & 0.99991 & 3.92 \\
4C +55.17 & 150.6 & 0.99990 & 3.89 & 94.1 & 0.99840 & 3.16 \\
Ton 116 & 148.8 & 0.96454 & 2.10 & 132.2 & 0.99941 & 3.44 \\
PG 1246+586 & 251.8 & 0.99983 & 3.76 & 198.1 & 0.99826 & 3.13 \\
RBS 1432 & 140.4 & 0.99798 & 3.09 & 111.0 & 0.99932 & 3.40 \\ \hline
\end{tabular}
}
\caption{The highest energy photons and their associated probability of originating from the proposed source.}
\label{T:HEPprob}
\end{table}

\begin{figure}
\centering
\includegraphics[width=0.5\textwidth]{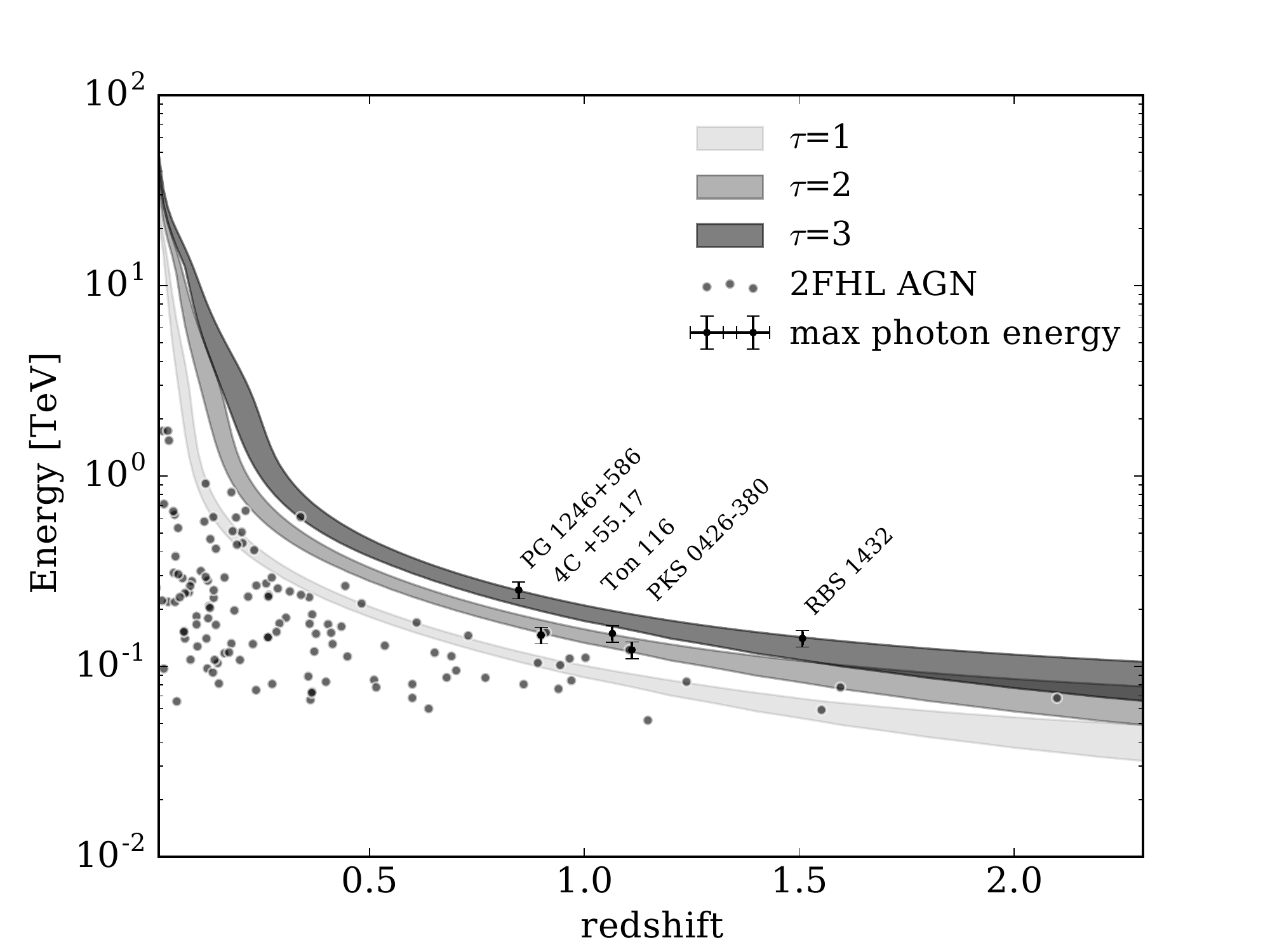}
\caption{Distribution of the detected energy compared to the redshift for all 2FHL sources with an available redshift, and the DBSCAN sources presented in this work. This is often referred to as the Fazio Stecker relation \citep{faziostecker}. Also shown are the different levels of EBL absorption, where the area covers the four different models used in this work.}
\label{F:fazsteck}
\end{figure}

As an initial consideration, the maximum energies found in Table \ref{T:HEPprob} and the redshift of the proposed sources were compared to objects in the rest of the 2FHL (where the 2FHL catalogue contains the maximum energy detected). This can be seen in Figure \ref{F:fazsteck} in which different levels of optical depth, starting at  $\tau=1$ (where the universe becomes optically thick to gamma-rays) have also been shown. What is immediately apparent is that the majority of 2FHL sources lie within, or close to, the optically thin ($\tau<1$) region, while the sources found in the DBSCAN VHE source sample are pushing out to larger values of $\tau$.

\begin{figure}
\centering
\includegraphics[width=0.5\textwidth]{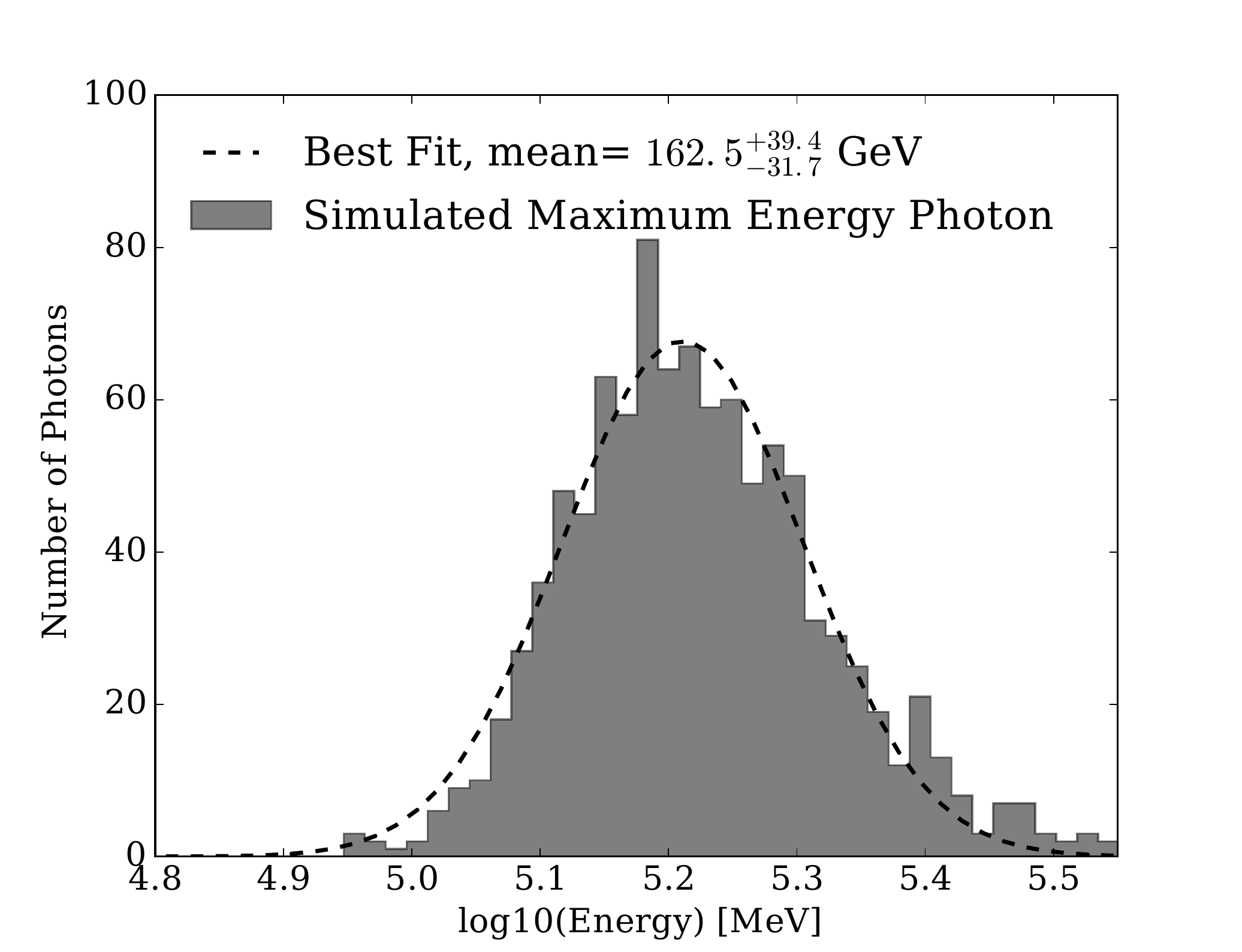}
\caption{Distribution of simulated maximum photon energy for source PG 1246+586.}
\label{F:maxEdist}
\end{figure}

Of the VHE sources, the largest derived optical depth corresponds to $\tau_{GIL12}=4.39$ for RBS 1432, indicating that a photon was detected despite 98.8\%  of the flux being attenuated. In order to evaluate the probability of observing the highest energy photons from each source, Monte Carlo simulations were performed as in \cite{fermiEBL}. Here, spectral parameters were drawn from a distribution based on the best fit intrinsic model and corresponding errors from section \ref{S:specAn}, including the spectral EBL absorption from the GIL12 model. Then using the \textit{Fermi} science tool \textsc{gtobsim},  eight year observations between 10~GeV and 1~TeV were simulated, taking into account the instrument response functions and spacecraft pointing history. Using 1000 simulations for each source, the number of observations of photons of a given energy was determined. An example of this for PG 1246+586 can be seen in Figure \ref{F:maxEdist}. 

From these simulations, it is evident that, in the 8 years of \textit{Fermi}-LAT observations, the highest energy photons detected are within expectation, with all but the 251~GeV photon from PG 1246+586 at 2.2$\sigma$ found at an energy less than one standard deviation from the simulated mean value. This result confirms that found in Section \ref{S:specAn}, that the observations presented in this work strengthen our confidence in the EBL models. 

Given that it has been shown that photons with energies greater than 100~GeV are expected from these sources, this also provides further evidence that they should be observable with ground based gamma-ray observatories which, with their much greater sensitivity, could provide stricter limits on EBL models. An important caveat here is that the analysis carried out in this work makes use of the near continuous \textit{Fermi}-LAT observations over 8 years. The assumption is therefore that these sources are non-variable over that time scale, or more importantly that the spectrum remains unchanged. The published variability index from the 3FGL suggests that this assumption is reasonable for all but PKS 0426-380 and RBS 1432, with the former exhibiting strong evidence for variability. Additionally, many of the sources from Section\ref{SS:highz2FHL} also show evidence of variability. It is expected that this may bias the results presented in this and the previous section. In a following paper, the effect of variability on the estimate of the EBL will be studied in detail.

\section{Discussion and Conclusion}
\label{S:conc}

\textit{Fermi}-LAT's excellent energy coverage enables it to be used to observe attenuation effects from the EBL. In this work, analysis of 16 high redshift sources has been presented, in which the EBL absorption is apparent. By modelling the energy spectrum from 100~MeV to around $\sim$10~GeV (redshift dependent), over which the effect of the EBL is negligible, it is possible to obtain an idea of the intrinsic spectrum of the source of interest.  Using this, the full data set from 100~MeV to 300~GeV and the absorption from a selection of EBL models, scaled by a correction factor $\alpha$, a measurement of the EBL scale was derived. By using a combination of power law and log parabola models, a best fit to the combined TS of each individual source revealed scaling factors of $\alpha_{\textrm{GIL12}}=0.95\pm0.05$, $\alpha_{\textrm{K\&D10}}=1.31\pm0.07$, $\alpha_{\textrm{FIN10}}=1.31\pm0.08$, $\alpha_{\textrm{FRA08}}=1.85\pm0.11$ and $\alpha_{\textrm{DOM11}}=1.85\pm0.11$ for the redshift range $0.897<z<1.596$. All models apart from GIL12 predict an EBL density less than suggested by this study. As the choice of spectral model introduces an unknown uncertainty, we also derived a conservative EBL scaling factor based solely on the log parabola model, From this it was found that $\alpha_{\textrm{GIL12}}=0.90\pm0.05$ $\alpha_{\textrm{K\&D10}}=1.24\pm0.07$, $\alpha_{\textrm{FIN10}}=1.24\pm0.08$, $\alpha_{\textrm{FRA08}}=1.71\pm0.11$ and $\alpha_{\textrm{DOM11}}=1.75\pm0.11$.

As mentioned previously, the source B0 218+357 has been observed using the ground-based MAGIC telescopes \citep{2016arXiv160901095M}. They used a combination of MAGIC and \textit{Fermi}-LAT data for a period of activity between the 11th of July 2014 and the 6th of August 2014, and modelling B0 218+357 with a power law, to obtain EBL correction factors for the following models\footnote{In the analysis of B0 218+357, a larger range of models were investigated. We present only their results for the models also considered in this work.}: $\alpha_{\textrm{FIN10}}=0.91 \pm 0.32_{stat} \pm 0.19_{sys}$, $\alpha_{\textrm{FRA08}}=1.19 \pm 0.42_{stat} \pm 0.25_{sys}$, $\alpha_{\textrm{DOM11}}=1.19 \pm 0.42_{stat} \pm 0.25_{sys}$ and $\alpha_\textrm{{GIL12}}=0.99 \pm 0.34_{stat}~^{+0.15 sys}_{-0.18 sys}$.

In addition to B0 218+357, the MAGIC collaboration used 11.8 hours of observations from a 2014 flare of 1ES 1011+496 ($z=0.212$) \citep{2016A&A...590A..24A} to obtain a scaling factor for the EBL models  $\alpha_{\textrm{DOM11}}=1.07~ (-0.20,+0.24)_{stat+sys}$ and $\alpha_{\textrm{FRA08}}=1.14~ (-0.14,+0.09)_{stat}$. This was followed by a study using 8 high frequency peaked BL Lac and 4 FSRQs within the redshift range  of $0.031<z<0.944$ (which also includes the observations of the two aforementioned sources) where a scaling factor of $\alpha_{\textrm{DOM11}}=0.99~(-0.56,+0.15)_{stat+sys}$ was derived \citep{2016arXiv161009633M}

Correction factors for the EBL have also been derived for less distant AGN by the HESS collaboration \citep{2013A&A...550A...4H}. Here data for 7 bright Blazars (Mrk 421 [\textit{z}=0.031], PKS 2005-489 [\textit{z}=0.071], PKS 2155-304 [\textit{z}=0.116], 1ES 0229+200 [\textit{z}=0.14], H 2356-309 [\textit{z}=0.165], 1ES 1101-232 [\textit{z}=0.186], and 1ES 0347-121 [\textit{z}=0.188]), over different periods of activity, were modelled in a similar way to that presented here. By fitting a range of different spectral models to the observed data, scanning the EBL correction parameter space, a value was obtained that maximises the likelihood. In the paper, only the \citet{FRAN08} model was considered, for which a correction factor of $\alpha_{\textrm{FRA08}}=1.27_{-0.15_{stat}}^{+0.18_{stat}} \pm 0.24_{sys}$ was obtained.

The most encompassing study using ground based observations was performed in \citet{2015ApJ...812...60B}. Here spectra from 38 sources in the redshift range $0.019<z<0.64$ were evaluated, using data from HESS, MAGIC, VERITAS, Whipple, ARGO-YBJ, TACTIC, HEGRA, Tibet and CAT. Using these data sets, along with local constraints taken from \citet{2013APh....43..112D}, the spectra were fitted with power law, log parabola and exponential cutoff powerlaw models, resulting in a derived EBL spectrum. Also presented were correction factors for current EBL models: 
$\alpha_{\textrm{GIL12}}=1.13\pm0.07$ 
$\alpha_{\textrm{K\&D10}}=1.52\pm0.14$, 
$\alpha_{\textrm{FIN10}}=1.48\pm0.07$, 
$\alpha_{\textrm{FRA08}}=1.05\pm0.07$ and 
$\alpha_{\textrm{DOM11}}=1.16\pm0.05$.

One of the main scientific objectives of the \textit{Fermi}-LAT instrument was to measure the effect of the EBL attenuation of high energy gamma-rays. In \citet{2012Sci...338.1190A}, data from the first 46 months of observation were used to model the EBL attenuation of 150 BL Lac type blazars. The method to determine the scaling of the EBL was the same as presented in this work, first modelling the intrinsic blazar spectrum and then fitting the higher energies, up to 500~GeV. In doing this, scaling factors for a large range of EBL models were determined. For the models used in this work, the corresponding weighted averages were found to be $\alpha_{\textrm{GIL12}}=0.67\pm0.14$ $\alpha_{\textrm{K\&D10}}=0.90\pm0.19$, $\alpha_{\textrm{FIN10}}=0.86\pm0.23$, $\alpha_{\textrm{FRA08}}=1.02\pm0.23$ and $\alpha_{\textrm{DOM11}}=1.02\pm0.23$. Also presented was the binned results for the \citet{FRAN08} model, giving $\alpha_{\textrm{FRA08}}=1.18^{+0.94}_{-0.81}$ ($z<0.2$), $\alpha_{\textrm{FRA08}}=0.82^{+0.41}_{-0.30}$ ($0.2<z<0.5$) and $\alpha_{\textrm{FRA08}}=1.29^{+0.43}_{-0.36}$ ($0.5<z<1.6$). 

\begin{figure*}
\includegraphics[width=\textwidth]{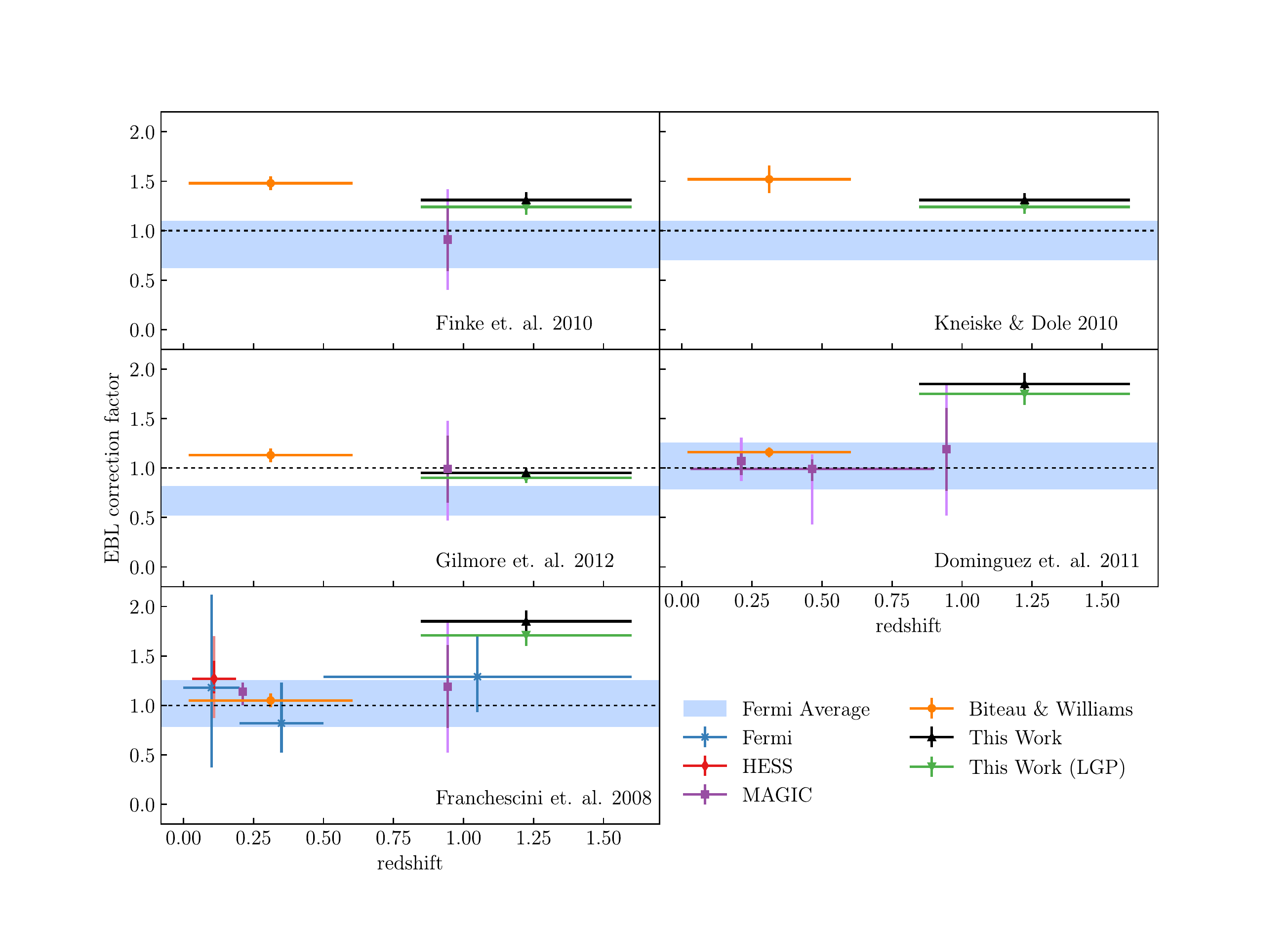}
\caption{Derived scaling factors for the EBL models considered in this work as a function of redshift. The HESS measurements are taken from  \citep{2013A&A...550A...4H} (red diamond), the MAGIC from  \citep{2016arXiv160901095M}, \citep{2016A&A...590A..24A} and \citep{2016arXiv161009633M} (purple square), the \textit{Fermi}-LAT points from \citep{2012Sci...338.1190A} and the combined ground based points from \citep{2015ApJ...812...60B} (orange circles). The black upward triangle is the value derived in this work using a combination of power law and log parabola models, whereas the green downward triangle shows the value derived when only considering log parabola models.}
\label{F:resfra}
\end{figure*}

The results from \citet{2016arXiv160901095M}, \citet{2016A&A...590A..24A}, \citet{2016arXiv161009633M}, \citet{2013A&A...550A...4H}, \citet{2012Sci...338.1190A} and \citet{2015ApJ...812...60B} are shown in comparison to the results derived in this paper in Figure \ref{F:resfra}. It can be seen that, for all but the GIL12, our results imply a larger density of the EBL at high redshifts. For all models we find a larger estimate than found by the \textit{Fermi}-LAT in \citet{2012Sci...338.1190A}; however they are consistent with the results found by the MAGIC collaboration \citep{2016arXiv160901095M}. When considering the binned \textit{Fermi}-LAT results for the FRA08 model, the last bin which encompasses the redshift range under investigation in this paper, is consistent with our results for log parabola models. When considering the EBL model types, it is suggested that the backward evolution models DOM11 and FRA08, which are based on the luminosity function of galaxies based on number counts, underestimate the EBL at large redshifts. This is not an unexpected result as these models match observed lower limits which are based on incomplete surveys. In a following paper we will attempt to investigate these trends in more detail, taking into account variability and source class.

% This implies there may be an increased effect of the EBL for high redshift when considering the \mbox{\citep{FRAN08}} model. In a following paper, this will be investigated further along with the effect of variability and source class.}

The second part of this work focused on  investigating the highest energy photons from the VHE sources found with DBSCAN. By considering the often used Fazio-Stecker relation, it seemed that these sources were pushing out to large optical depths. Upon further investigation using Monte Carlo simulations, it was shown that the observed photon energies were not unexpected (less than 1 standard deviation from the mean energy or $2.2\sigma$ for PG 1246+586), illustrating the pitfalls of relying solely on the Fazio-Stecker relation.

Given the expected performance of the future ground based observatory, CTA, which will be able to place even stronger constraints on the EBL, it was shown that several of the sources presented in the work should be observable by CTA (See Figure \ref{F:sed1} and \ref{F:sed2}). Indeed, the observation of VHE emission from PKS 0426-380, 4C +55.17, Ton 116, PG 1246+586 and RBS 1432 is expected, providing further evidence that these sources should be observable with CTA.

\section*{Acknowledgments}

 TPA would like to acknowledge the support from the UK Science and Technology Facilities Council grant ST/K501979/1. AMB would like to acknowledge the financial support of Durham University. This work has made use of publicly available \textit{Fermi}-LAT data from the High Energy Astrophysics Science Archive Research Center (HEASARC), provided by NASAs Goddard Space Flight Center. Finally, we would like to thank the reviewer for their insightful comments which helped improve the quality of this work.

\bibliography{main}

\appendix
\section{Detailed Source Analysis}
\label{A:1}

\begin{figure*}
\centering
\begin{tabular}{cc}
\includegraphics[width=0.5\textwidth]{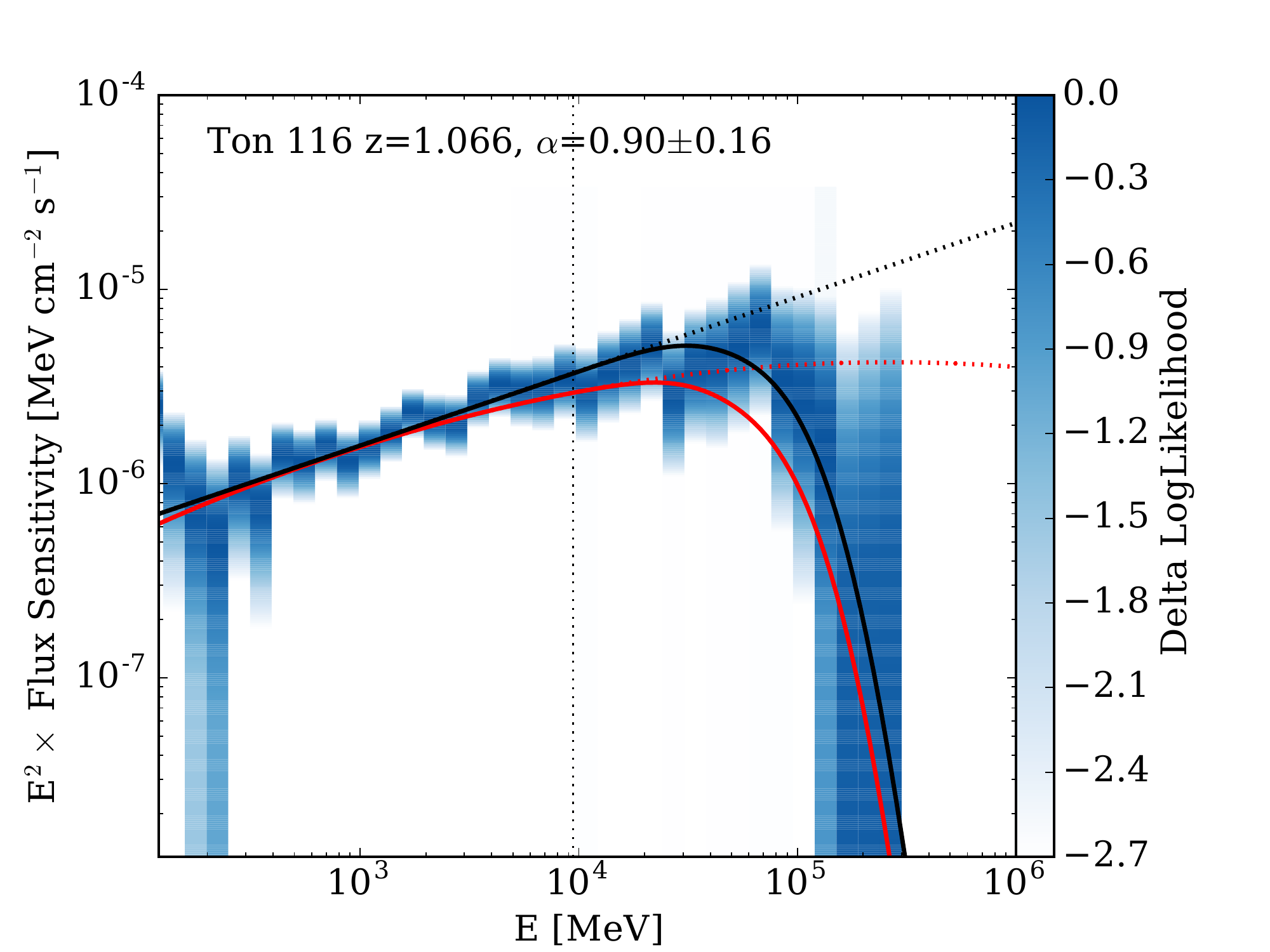} &
\includegraphics[width=0.5\textwidth]{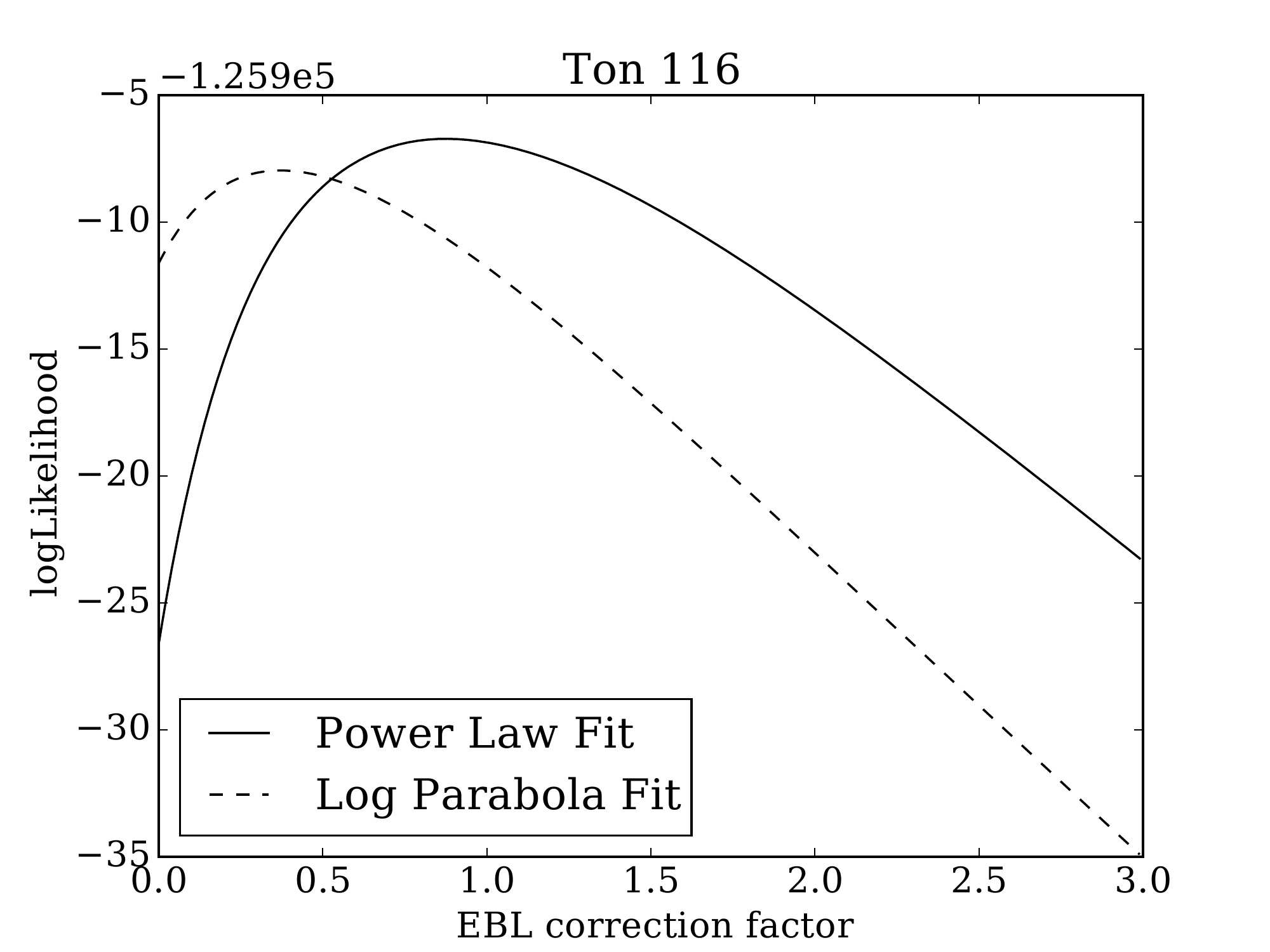} \\
a) & b) \\
\end{tabular}
\caption{a) The SED for Ton 116 showing the power law spectral model (black) and the log parabola model (red). Each of these models is also shown convolved with the EBL attenuation from the GIL12 model (dotted lines).  b) The likelihood distribution for each spectral model as a function of the EBL scaling factor. The power law model is taken as the preferred model in the case of Ton 116.}
\label{F:A1}
\end{figure*}

To provide further clarity of the method by which the EBL scaling factor was determined, a more detailed description is given here. For simplicity, the source Ton 116 (\textit{z}=1.066) is used as an example throughout.

For each source, an initial fit was obtained from 100~MeV to an energy where the EBL absorption is less than 0.1\%. For Ton 116, at a redshift of $z=1.066$, this corresponds to an energy of 9.24~GeV using the GIL12 model. The source of interest in the model file is then replaced with a file function based on the intrinsic spectral model convolved with the EBL absorption including the scaling factor $\alpha$: 

\begin{equation}
\frac{dN}{dE}_{obs} = e^{-\alpha \cdot \tau (E,z,n)} \cdot \frac{dN}{dE}_{int}.
\end{equation}

A series of model files were then created using the scaled spectral model for $\alpha$ values in the range 0 to 3 at 0.01 intervals. The sources within the ROI and the background models are fixed to the initial fit results and a second fit is applied, returning the log likelihood for the entire ROI. In the left panel of Figure \ref{F:A1} the intrinsic power law and log parabola spectral models are compared to both the data and the absorbed spectral models ($\alpha=1$ for the GIL12 model), in the right panel the resulting log likelihood for each $\alpha$ value is shown. The standard method for choosing between models, as specified by the \textit{Fermi} collaboration, is to calculate the Test Statistic (TS) that the proposed model is preferred over the power law model, where the TS is defined as

\begin{equation}
\textrm{TS}=2\textrm{log}[\mathcal{L}_{\textrm{max}}(Log Parabola)/\mathcal{L}_{\textrm{max}}(Power Law)].
\end{equation}

For Ton 116, the TS value calculated is 1.24 and therefore the default power law model is retained (a TS$>$16 is required to justify the choice of an alternative model). Following this, the likelihood distribution is converted into a TS distribution, where this is now the TS of a given $\alpha$ value over the case where there is no EBL absorption ($\alpha=0$), i.e. 

\begin{equation}
\textrm{TS}=2\textrm{log}[\mathcal{L}(\alpha)/\mathcal{L}(\alpha=0)].
\end{equation}

Due to the asymmetrical nature of the likelihood/TS distribution, the mean and variance were calculated as in \cite{2004physics...3086D}, where the pdf is derived from the TS distribution by calculating $e^{TS(\alpha)}$ which is normalised based on the total area. From this the expected value is derived as 

\begin{equation}
\bar{\alpha} = \int \textrm{pdf}(\alpha) \cdot \alpha~ d\alpha
\end{equation}

and the variance as

\begin{equation}
Var = \int (\alpha - \bar{\alpha})^2 \cdot \textrm{pdf}(\alpha)~d\alpha.
\end{equation}

For Ton 116, this corresponds to $\alpha=0.90\pm0.16$. Lastly, in order to determine an overall EBL scaling factor, the TS distributions for each source are summed to obtain an combined TS where the expected value and variance are calculated as before (See Figure \ref{F:alphascan}).

\label{lastpage}

\end{document}